\documentclass[%
 reprint,
 amsmath,amssymb,
 aps,
pra,
]{revtex4-1}

\usepackage[ansinew]{inputenc}
\usepackage{graphicx}
\usepackage{dcolumn}
\usepackage{bm}
\usepackage{pst-all}
\usepackage{float}

\begin{document}

\preprint{APS/123-QED}

\title{Investigation of the $\Lambda(1405)$ line shape in pp collisions}
\author{ J.~Siebenson$^{1}$}
\author{L.~Fabbietti$^{1}$}

\affiliation{
\mbox{$^{1}$Excellence Cluster 'Origin and Structure of the Universe', Technische Universit\"at M\"unchen , 85748~Garching, Germany}\\
} 






\begin{abstract}

In this work we investigate different possible interpretations of the  $\Lambda(1405)$ signal associated with the production of the $\Lambda(1405)$ in $p+p$ reactions at 3.5~GeV beam kinetic energy measured by the HADES collaboration. We study the influence of interference effects between the $\Lambda(1405)$ resonance and the non$\--$resonant background. The two poles nature of the $\Lambda(1405)$, which is supported by most of the theoretical models,  is also discussed with emphasis on the relative contributions of the two complex poles to the formation of the resonance in $p+p$ reactions.
  
\end{abstract}

\pacs{Valid PACS appear here}
\maketitle


\section{Introduction} 
\begin{figure}[tbp]
	\centering
		\includegraphics[width=0.45\textwidth]{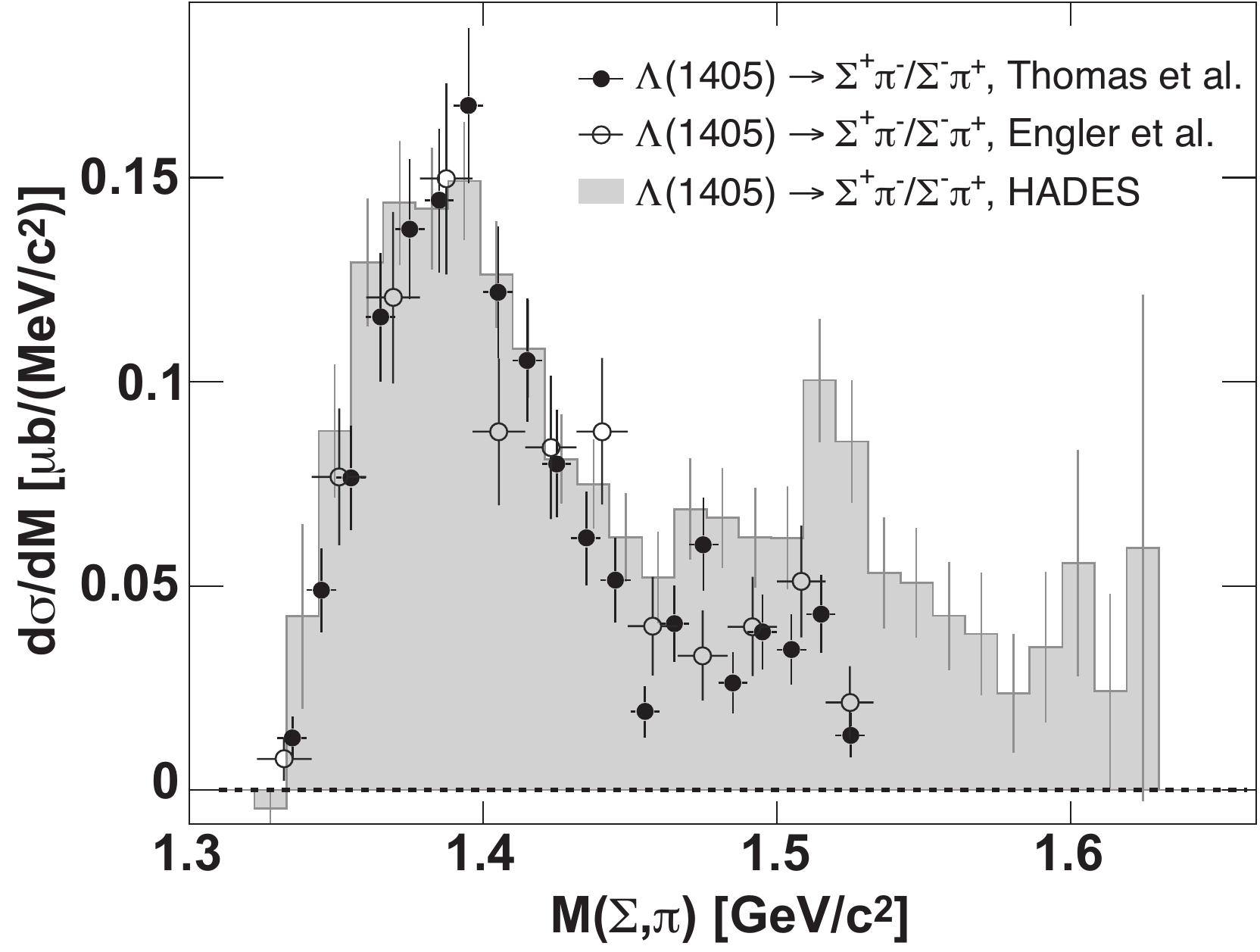}
	\caption{Comparison between the HADES data and the data of \cite{Thomas:1973uh} and \cite{Engler:1965zz}, measured in $\pi^-+p$ reactions. All three spectra show the sum of the $\Sigma^+\pi^-$ and $\Sigma^-\pi^+$ data samples.}
	\label{fig:LA1405_HADES_Thomas_Comp}
\end{figure}

Recently, the HADES collaboration has measured $\Lambda(1405)$ production in proton-proton reactions at a beam kinetic energy of 3.5~GeV \cite{Agakishiev:2012xk}, where the $\Lambda(1405)$ hyperon has been reconstructed in the charged $\Sigma^{\pm}\pi^{\mp}$ decay channels. By investigating the $\Sigma\pi$ invariant mass distributions, clear peak structures below 1400~MeV/$c^2$ were observed. These structures were interpreted by a small contribution of $\Sigma(1385)^0$ and a large contribution of a low mass $\Lambda(1405)$ signal (see also \cite{Agakishiev:2012ja}). Besides this, the spectra also showed a considerable contribution by $\Lambda(1520)$ production and by non$\--$resonant $\Sigma\pi$ production, resulting in a phase space like background below the $\Lambda(1405)$. The experimental data were finally described by an incoherent sum of Monte Carlo simulations, where the $\Lambda(1405)$ was simulated to follow a Breit-Wigner type distribution with a Breit-Wigner mass of 1385~MeV/$c^2$ and a width of 50~MeV/$c^2$. With help of these simulations the experimental data were corrected for the effects of acceptance and efficiency. These corrected data allow to compare any (more advanced) model to the obtained $\Sigma\pi$ invariant mass distributions. \\
The experimental data, where the maximum of the $\Sigma\pi$ missing mass (see Fig.~\ref{fig:LA1405_HADES_Thomas_Comp}) lies below the nominal value of $1405\,\mathrm{MeV/}c^2$ associated to the $\Lambda(1405)$, suggest a shift of this resonance towards lower masses.
In this paper we want to address different possible explanations for the observed mass shift of the $\Lambda(1405)$ in the new HADES data and aim to stimulate theorists to further investigate the production of this resonance in $p+p$ reactions.
The $\Lambda(1405)$ spectral function measured in $p+p$ collisions by the HADES collaboration differs from the predictions by theoretical models and also from some of the experimental observations. These differences reside in the complex nature of the resonance.

 From the theoretical point of view, the $\Lambda(1405)$ is normally treated in a unitarized coupled channel framework based on chiral SU(3) dynamics \cite{Hyodo:2007jq,Borasoy:2006sr,Kaiser:1995eg,Oset:1997it}, where this resonance is generated dynamically as a coherent sum of two different poles. The first pole, $z_1$, is located at higher energies of around 1420~MeV and is mainly associated with a narrow quasi-bound $\bar{K}N$ state. The second pole, $z_2$, is found at lower energies of about 1390~MeV and this pole strongly couples to a broad $\Sigma\pi$ resonance. As the relative contribution of these two states depends on the entrance channel, also the observed properties of the $\Lambda(1405)$ could differ for different reactions. Therefore, in order to understand the complex formation process of the $\Lambda(1405)$, it is important that experiments measure this resonance in different collision systems, and that, at the same time, theory provides appropriate models for each of those systems. 
 
First we refer to the measurement presented by the ANKE collaboration in \cite{Zychor:2007gf}, where the $\Lambda(1405)$ spectral shape is reconstructed in $p+p$ collisions at 2.83~GeV beam kinetic energy out of the decay into $\Sigma^0$ and $\pi^0$ pairs.
These data are within the systematic errors and the statistical significance of the mass bins around 1400~MeV/$c^2$ consistent with the HADES results.\\
From the theory side, the authors of \cite{Geng:2007vm} followed a unitarized coupled channel approach based on the chiral Lagrangian in order to 
predict the $\Lambda(1405)$ line shape in $p+p$ reactions for the $\Sigma^0\pi^0$ decay channel. In their Ansatz the $\Lambda(1405)$ was 
generated from pion, kaon and $\rho$ meson exchange mechanisms, all of them leading to a different coupling to the two $\Lambda(1405)$ poles. The 
coherent sum of all contributions results in a $\Lambda(1405)$ line shape with a maximum in the $\Sigma\pi$ mass distribution at around 
$1410\,\mathrm{MeV/}c^2$. With this approach the authors of \cite{Geng:2007vm} delivered a result compatible with the $\Lambda(1405)$ signal, measured by the ANKE collaboration in the $\Sigma^0\pi^0$ decay channel \cite{Zychor:2007gf}. 
This calculation is the only one available for $p+p$ reactions but since the HADES data refer to the charged decay channels
a quantitative comparison results difficult. Nevertheless, the results by \cite{Geng:2007vm} have been used in this work as a starting point to model the $\Lambda(1405)$ as the combination of two Breit-Wigner functions.\\
In general, the HADES data show a larger contribution by the non$\--$resonant $\Sigma\pi$ production in comparison with ANKE. In particular, it has 
been shown in \cite{Agakishiev:2012qx} that the $\Delta^{++}$ is strongly coupling to the $\Sigma^- \pi^+ p K^+$ final state via the reaction $p+p
\rightarrow \Sigma^- + \Delta^{++} + K^+$ and it cannot be excluded that $N^*/\Delta^0$ resonances contribute to the $\Sigma^+\pi^-pK^+$ channel
 as well. These contributions might appear in the $I=0$ channels and hence interfere with the resonant amplitude.

Aside from the predictions by models that consider the $\Lambda(1405)$ as the combination of two main poles, one has to consider those models where 
a single pole is associated to the formation of the resonance. 
In \cite{Hassanvand:2012dn} a phenomenological  ${\mathrm{\bar{K}}p}$ interaction was employed to derive the mass distribution of $
\Lambda(1405)$.
This approach was used to fit the HADES data. For this purpose all simulated contributions, mentioned in
\cite{Agakishiev:2012xk}, were subtracted so that only the pure $\Lambda(1405)$ signal was left. The fit within this phenomenological approach
 results into a good description of the experimental data and allowed to extract the mass and width of the $\Lambda(1405)$ to
$1405^{+11} _{-9}\mathrm{ MeV/c^2}$ and $62\pm 10 \mathrm{MeV/}c^2$, which is in good agreement with the PDG values \cite{PDG} and
does not imply any evident shift of the spectral function.

In the sector of $\gamma$-induced reactions, the CLAS collaboration has recently published new high quality data on the $\Lambda(1405)$ \cite{Moriya:2013eb,Moriya:2013hwg,Schumacher:2013vma}. In these reports, all three $\Sigma\pi$ decay channels have been investigated simultaneously for different incident photon energies. The observed $\Lambda(1405)$ spectral shape partially appears at higher masses, clearly above 1400 MeV/$c^2$. 
The structures measured by CLAS show a dependency on the photon incident energy and also differ among the three $\Sigma\pi$ decay channels. 
Indeed, one has to consider that for photon-induced reactions the interference between the $I=0$ and $I=1$  channels is not negligible as for proton- and pion-induced reactions. 
Recent theoretical works \cite{PhysRevC.87.055201,Roca:2013cca} employ parameters fitted to the CLAS experimental data which 
allow for a small SU(3) breaking. This study shows that more precise calculations including higher order corrections could be needed in this sector 
and also suggests the possible existence of a $I=1$  bound state in the vicinity of the  
$\bar{\mathrm{K}N}$ threshold. Additionally, the CLAS collaboration reported recently on the first observation of the $\Lambda(1405)$ in electron-
induced reactions \cite{Lu:2013nza} showing very different features respect to the photon-induced results. 

In the sector of pion-induced reactions, Thomas et al. \cite{Thomas:1973uh} and Engler et al. \cite{Engler:1965zz} have measured $\pi^-+p$ 
collisions at a beam momentum of 1.69~GeV/$c$ and have reconstructed the $\Lambda(1405)$ from its decay into $\Sigma^{\pm}\pi^{\mp}$. The 
results for the $\Sigma^+\pi^-+\Sigma^-\pi^+$ invariant mass spectra are shown in the black and open data points in Fig.~
\ref{fig:LA1405_HADES_Thomas_Comp}, respectively.
According to \cite{Thomas:1973uh}, the spectra consist of several contributions, namely 46\% $\Lambda(1405)$, 8\% $\Sigma(1385)^0$, 3\% $\Lambda(1520)$ and 43\% non$\--$resonant $\Sigma\pi$ production. The broad peak structure around 1400~MeV/$c^2$ is mainly identified with the $\Lambda(1405)$ signal. 
This experimental spectrum, however, is not fully understood from the theoretical side, which expects a large contribution from the first $\Lambda(1405)$ pole, shifting the expected $\Lambda(1405)$ distribution to higher mass values \cite{Hyodo:2003jw}. Therefore, the question arises if, in case of $\pi^-$-induced reactions, the coupling to the second, broad pole at $\approx 1390$~MeV is underestimated by theory.\\
It is interesting to compare the results from $\pi^-$-induced reactions to the new HADES data. The gray histogram in Fig.~\ref{fig:LA1405_HADES_Thomas_Comp} shows the summed invariant mass spectrum of $\Sigma^+\pi^-+\Sigma^-\pi^+$ of \cite{Agakishiev:2012xk}. As the relative contributions from $\Lambda(1405)$, $\Sigma(1385)^0$, $\Lambda(1520)$ and non$\--$resonant $\Sigma\pi$ channels in these data are quite similar to the ones in the considered $\pi^-+p$ reactions, a comparison between the data sets is justified. In order to allow such a comparison, the data from \cite{Thomas:1973uh} and \cite{Engler:1965zz} have been scaled appropriately. The agreement between the three spectra in the region around 1400~MeV/$c^2$ is excellent, indicating that all measurements observe a similar low mass $\Lambda(1405)$ signal. Furthermore, this suggests that in both, $p+p$ and $\pi^-+p$ reactions, the broad $\Sigma\pi$ pole might be dominant in the coupling to the $\Lambda(1405)$ state. 
\noindent However, since the measured $\Sigma\pi$ final state does not contain the pure signature of the $\Lambda(1405)$, but also contains a 
considerable contribution of non$\--$resonant background, the interpretation of this result is not straight forward.
 \section{Influence of interference effects}
\begin{figure}[tbp]
	\centering
		\includegraphics[width=0.45\textwidth]{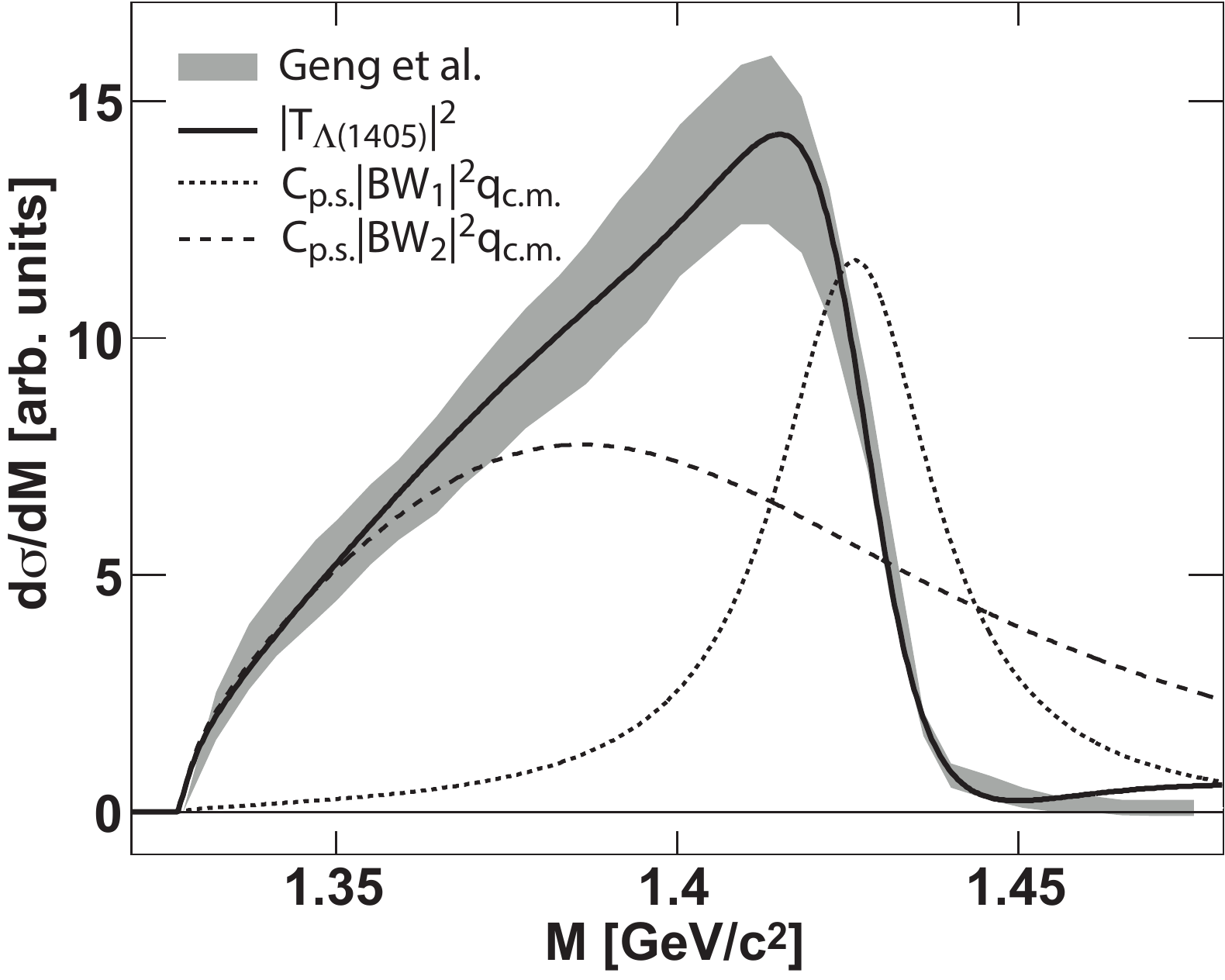}
	\caption{The $\Lambda(1405)$ spectral shape calculated in \cite{Geng:2007vm} (gray histogram). The spectrum is fitted with Eq.~(\ref{equ:DoubleBW}). The two phase space modified Breit-Wigner functions $C_{\mathrm{p.s.}}(m)\left|BW_1(m)\right|^2q_{\mathrm{c.m.}}$ and $C_{\mathrm{p.s.}}(m)\left|BW_2(m)\right|^2q_{\mathrm{c.m.}}$ are shown in the dotted and dashed lines.}
	\label{fig:OsetFitted}
\end{figure}
\begin{figure}[tbp]
	\centering
		\includegraphics[width=0.46\textwidth]{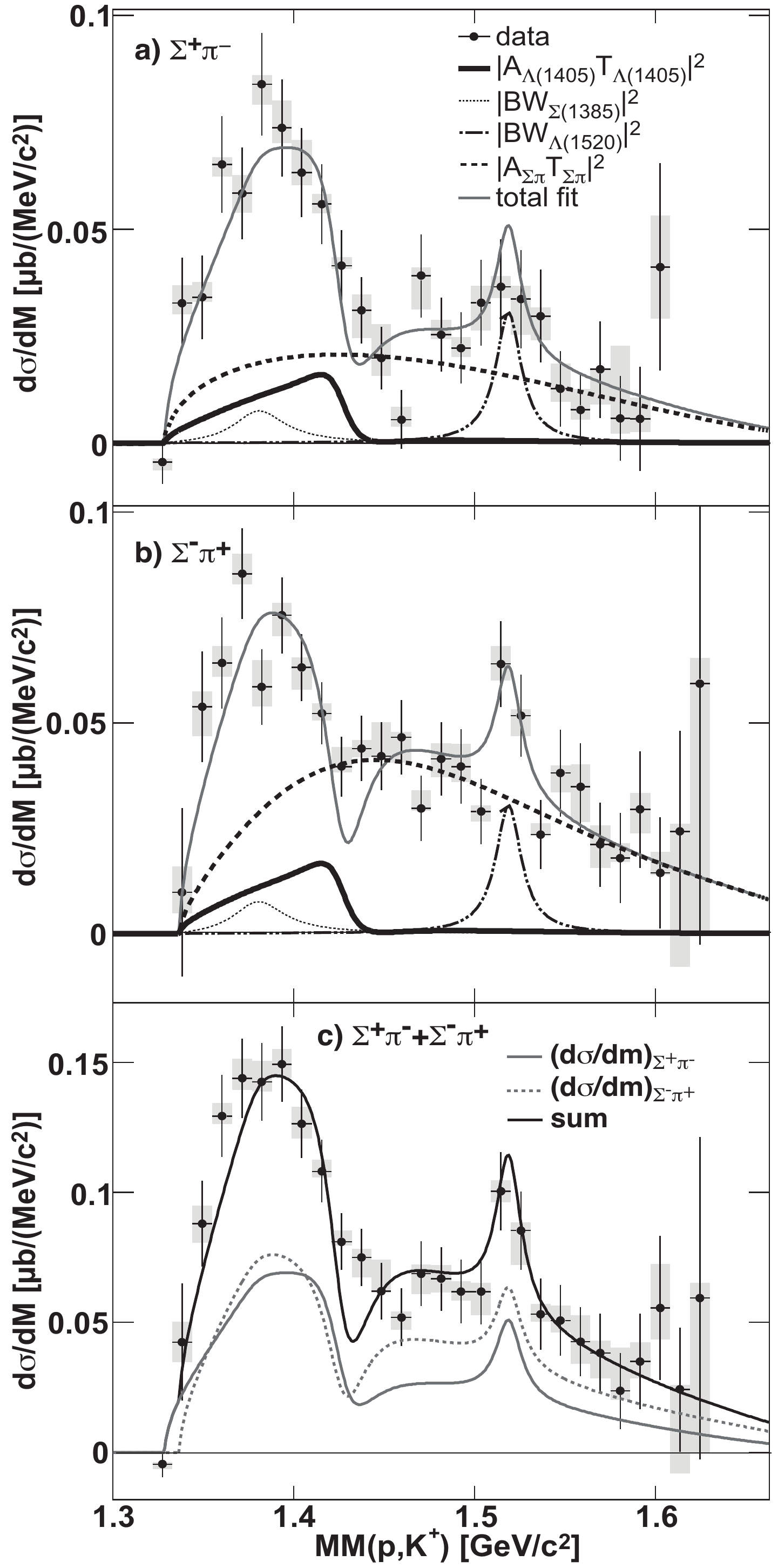}
	\caption{ Missing mass spectrum to proton and $K^+$. The black data points are the measurements of \cite{Agakishiev:2012xk}, for the $\Lambda(1405)$ in the $\Sigma^+\pi^-$ (a) and $\Sigma^-\pi^+$ (b) decay channel. Panel c) shows the summed spectrum of a) and b). The black lines in a) and b) are the results from the simultaneous fit with Eq.~(\ref{equ:DistSpPm}) and (\ref{equ:DistSmPp}), the gray line represents the sum of all fitted functions. In c) the fit functions corresponding to the  $\Sigma^+\pi^-$ and $\Sigma^-\pi^+$ channels respectively are shown in gray and the sum of both functions is shown in the black line.}
	\label{fig:Interferences}
\end{figure}

  
Interference between the resonant and the non$\--$resonant amplitudes can affect the observed mass distribution considerably. In order to evaluate this scenario, we take the result of the chiral Ansatz \cite{Geng:2007vm} as a starting point to develop a simple model, which we use to parametrize the $\Lambda(1405)$ amplitude and to interpret the HADES data. The predicted $\Sigma\pi$ spectrum of \cite{Geng:2007vm} is shown in the gray band of Fig.~\ref{fig:OsetFitted}, with a peak at 1410~MeV and with the typical $\Lambda(1405)$ shape, having a sharp drop to the $\bar{K}N$ threshold. The idea is to reconstruct  this spectrum by a coherent sum of two Breit-Wigner functions (BW), where each of these BW amplitudes represents one of the two $\Lambda(1405)$ poles. This approach is similar to what has been proposed in \cite{Jido:2003cb} to represent via Breit-Wigner distributions the results of the unitarized coupled channel calculations. It is clear that any variation of the parameters associated to the two poles does not fulfill unitarity anymore and hence our procedure is not equivalent to a full-fledged calculation. Still it is interesting to see how much the parameters have to be modified to fit the experimental data.
 Within this approach the  total $\Lambda(1405)$ amplitude reads then as follows:
\begin{eqnarray}
\frac{d \sigma}{d m}&=&\left|T_{\Lambda(1405)}\right|^2= \nonumber \\
&=&C_{\mathrm{p.s.}}(m)\left|BW_1(m)e^{i\varphi_1}+BW_2(m) \right|^2q_{\mathrm{c.m.}} \label{equ:DoubleBW} \\
&&\mbox{with } BW_i=A_i\frac{1}{\left(m-m_{0,i}\right)^2+im_{0,i}\Gamma_{0,i}} \nonumber
\end{eqnarray}
$C_{\mathrm{p.s.}}(m)$ is a dimensionless weight function, normalized to unity in the mass range of 1280-1730~MeV/$c^2$. This function considers the limited production phase space of the $\Lambda(1405)$ in $p+p$ reactions. $q_{\mathrm{c.m.}}$ is the decay momentum of $\Sigma$ and $\pi$ in the $\Lambda(1405)$ rest frame (in units of [MeV/c]). The Breit-Wigner function is a simple relativistic parametrization with amplitude $A_i$ in units of $\left[\sqrt{\mu b/\mbox{c}}\cdot \mbox{MeV/c}^2\right]$, mass $m_{0,i}$ and width $\Gamma_{0,i}$ in units of [MeV/$c^2$]. Thus, the whole expression has dimensions of $\left[\frac{\mu b}{\mbox{MeV/c}^2}\right]$. We also introduce a free phase $e^{i\varphi_1}$, which determines the interference between the two Breit-Wigner functions. Furthermore, we make use of the recent coupled channel calculations by Ikeda et al. \cite{Ikeda:2012au}, which are constrained by the new SIDDARTHA data on kaonic hydrogen \cite{Bazzi:2011zj}. In this work the $\mathrm{\bar{K}N}$ pole was found at $z_1=1424^{+7}_{-23}+i26^{+3}_{-14}$~MeV, while the $\Sigma\pi$ pole was found at $z_2=1381^{+18}_{-6}+i81^{+19}_{-8}$~MeV. With these values we constrain the Breit-Wigner mass $m_ic^2=Re(z_i)$ and the Breit-Wigner width $\Gamma_{0,i}c^2=2Im(z_i)$ of our model to vary only within the given ranges.\\
Although the parametrization of Eq.~(\ref{equ:DoubleBW}) is simplified compared to the advanced calculations in \cite{Geng:2007vm}, it still allows us to reconstruct the spectral shape in Fig.~\ref{fig:OsetFitted} (gray band).
By fitting the Eq.~(\ref{equ:DoubleBW}) to the theoretical prediction,  we obtain the black distribution with the fit parameters listed in Table~\ref{tab:tableInter}.
\begin{table}[tbp]
\caption{\label{tab:tableInter}%
Table with fit parameters obtained by fitting Eq.~(\ref{equ:DoubleBW}) to the theoretical prediction of \cite{Geng:2007vm} shown in Fig.~\ref{fig:OsetFitted}.}
\begin{ruledtabular}
\begin{tabular}{cccccc}
\textrm{$m_1$}&
\textrm{$\Gamma_1$}&
\textrm{$m_2$}&
\textrm{$\Gamma_2$}&
\textrm{$A_1/A_2$}&
\textrm{$\varphi_1$}\\
\colrule
1426 & 28 & 1375 & 147 & 0.23 & 205$^{\circ}$  \\
\end{tabular}
\end{ruledtabular}
\end{table} 
This distribution is consistent with the gray band in Fig.~\ref{fig:OsetFitted} . Especially the peak structure around 1410~MeV/$c^2$ and the drop to the $\bar{K}N$ threshold is reproduced correctly. Additionally included in the figure are the absolute contributions of the two Breit-Wigner functions $C_{p.s.}(m)\left|BW_1(m)\right|^2q_{c.m.}$ and $C_{p.s.}(m)\left|BW_2(m)\right|^2q_{c.m.}$ (dotted and dashed lines).\\ 
In this way we have fixed the parametrization of the $\Lambda(1405)$ and can now study the maximal interference effects with the non$\--$resonant background. 
For this purpose, we fit the HADES results for the $\Sigma^+\pi^-$ and $\Sigma^-\pi^+$ invariant mass distributions simultaneously with the following two functions:
\begin{widetext}
\begin{eqnarray}
\left(\frac{d\sigma}{dm}\right)_{\Sigma^+\pi^-}=\left|A_{\Lambda(1405)}T_{\Lambda(1405)}+e^{i\alpha}A_{\Sigma^+\pi^-}T_{\Sigma^+\pi^-}\right|^2+\left|BW_{\Sigma(1385)^0}\right|^2 + \left|BW_{\Lambda(1520)}\right|^2 \label{equ:DistSpPm} \\
\left(\frac{d\sigma}{dm}\right)_{\Sigma^-\pi^+}=\left|A_{\Lambda(1405)}T_{\Lambda(1405)}+e^{i\beta}A_{\Sigma^-\pi^+}T_{\Sigma^-\pi^+}\right|^2+\left|BW_{\Sigma(1385)^0}\right|^2 + \left|BW_{\Lambda(1520)}\right|^2 \label{equ:DistSmPp}
\end{eqnarray} 
\end{widetext}
The contributions from $\Sigma(1385)^0$ and $\Lambda(1520)$ are parameterized as Breit-Wigner functions so that they match the extracted shapes and yields reported in \cite{Agakishiev:2012xk}. We assume here that they do not interfere with the other contributions to the $\Sigma\pi$ invariant mass spectra and thus add them incoherently in the Eq.s~(\ref{equ:DistSpPm}) and (\ref{equ:DistSmPp}). \\
The $\Lambda(1405)$ is parameterized as described above. $A_{\Lambda(1405)}$ is a free fit parameter which determines the absolute yield of $\Lambda(1405)$.\\      
The non$\--$resonant background shapes for the $\Sigma^+\pi^-$ and $\Sigma^-\pi^+$ channels ($T_{\Sigma^+\pi^-}$ and $T_{\Sigma^-\pi^+}$) are described by modified polynomials of 4th order, which read as follows: 
\begin{eqnarray}
T_{\Sigma\pi}(m)=\left[C_{p.s.}(m)q_{c.m.}\sum^{4}_{n=0}a_nm^n\right]^{\frac{1}{2}} \label{equ:NonRes}
\end{eqnarray}
This parametrization has no physical meaning but it was chosen such to describes the simulated $\Sigma\pi$ invariant mass distributions in \cite{Agakishiev:2012xk}. The parameters $a_n$ are given in units of $\left[\left(\mbox{MeV/c}^2\right)^{-n}\right]$ and their values are listed in Table~\ref{tab:table1}.
\begin{table}[tbp]
\caption{\label{tab:table1}%
Table with coefficients for the description of the non$\--$resonant background according to Eq.~(\ref{equ:NonRes}).}
\begin{ruledtabular}
\begin{tabular}{cccc}
\textrm{$T_{\Sigma\pi}$}&
\textrm{$a_0$}&
\textrm{$a_1$}&
\textrm{$a_2$}\\
\colrule
$T_{\Sigma^+\pi^-}$ & $7.949\cdot10^{-3}$ & $-4.412\cdot10^{-7}$ & $-1.558\cdot10^{-9}$    \\
$T_{\Sigma^-\pi^+}$ & $-2.387\cdot10^{0}$ & $2.512\cdot10^{-3}$ & $1.315\cdot10^{-6}$   \\
\colrule\\
\textrm{$T_{\Sigma\pi}$}&
\textrm{$a_3$}&
\textrm{$a_4$}\\
\colrule
$T_{\Sigma^+\pi^-}$ & $2.942\cdot10^{-12}$ & $-1.121\cdot10^{-15}$ \\
$T_{\Sigma^-\pi^+}$ & $-2.254\cdot10^{-9}$ & $6.481\cdot10^{-13}$ \\
\end{tabular}
\end{ruledtabular}
\end{table}
The modified polynomials are multiplied by constant factors $A_{\Sigma^+\pi^-}$ and $A_{\Sigma^-\pi^+}$ (see~Eq. (\ref{equ:DistSpPm}) and (\ref{equ:DistSmPp})), which have dimensions of $\left[\frac{\sqrt{\mu b/\mbox{c}}}{\mbox{MeV/c}^2}\right]$. These are again free fit parameters, determining the absolute yields of the non$\--$resonant channels, where a value of $A_{\Sigma\pi}=1$ corresponds to the yield extracted in \cite{Agakishiev:2012xk}. Furthermore, complex phases $e^{i\alpha}$ and $e^{i\beta}$ have been included so that the modeled background can interfere with the $\Lambda(1405)$ amplitude. The values of these phases are determined by the fitting procedure as well. Hence, the simultaneous fit of the two functions (\ref{equ:DistSpPm}) and (\ref{equ:DistSmPp}) to the experimentally determined $\Sigma^+\pi^-$ and $\Sigma^-\pi^+$ invariant mass distributions is characterized by five free parameters.\\
We consider here a scenario of maximal interference, which means that the whole non$\--$resonant background interferes with the $\Lambda(1405)$ amplitude.
 The best results of the fitting procedure (gray lines) are illustrated together with the experimental data in Fig.~\ref{fig:Interferences} panel a) and b). The black lines show the amplitude of the $\Lambda(1405)$, the red lines the contribution by the non$\--$resonant $\Sigma\pi$ channels and the gray lines correspond to the coherent sum of all contributions. The fit parameters are listed in Table~\ref{tab:table2}.  
\begin{table}[tbp]
\caption{\label{tab:table2}%
Obtained fit parameters after fitting Eq.s~(\ref{equ:DistSpPm}) and (\ref{equ:DistSmPp}) to the experimental data points in Fig.~\ref{fig:Interferences}.}
\begin{ruledtabular}
\begin{tabular}{ccccc}
\textrm{$A_{\Lambda(1405)}$}&
\textrm{$A_{\Sigma^+\pi^-}$}&
\textrm{$A_{\Sigma^-\pi^+}$}&
\textrm{$\alpha$}&
\textrm{$\beta$}\\
\colrule
$1.06$ & $0.93$ & $1.04$ & $67^{\circ}$ & $109^{\circ}$  \\
\end{tabular}
\end{ruledtabular}
\end{table}
A reasonable description of the data is achieved, expressed in a normalized $\chi^2$ value of 1.23. In panel c) of Fig.~\ref{fig:Interferences} we show the sum of the two invariant mass distributions and compare it to the sum of our model (black line). \\
The main message of the three pictures is that, assuming a maximal interference between the $\Lambda(1405)$ and the non$\--$resonant contributions and the large phases given in Table~\ref{tab:table2}, one obtains a shift of the maximum to lower masses. This is the case although the mass maximum of the initial $\Lambda(1405)$ amplitude is located around 1400~MeV/$c^2$. By integrating $\left|A_{\Lambda(1405)}T_{\Lambda(1405)}\right|^2$ (black lines in panel a) and b)), the total cross section of the reaction $p+p\rightarrow\Lambda(1405)+p+K^+$ is determined to 3.3 $\mu b$. This is considerably smaller than the value extracted in \cite{Agakishiev:2012xk}, where an incoherent approach was used with a low mass $\Lambda(1405)$ signal.\\    
At this point one has to emphasize some details of the model used in this work. First, the parametrization of the $\Lambda(1405)$ as a simple sum of two Breit-Wigner amplitudes is not equivalent to the full coupled channel calculation of \cite{Geng:2007vm}. A second point is the description of the non$\--$resonant background $T_{\Sigma\pi}$ as polynomial functions. This is certainly a simplification. Also the assumption that the non$\--$resonant channels just have a constant complex phase is oversimplified.
Furthermore, we assume that the whole $\Sigma\pi$ non$\--$resonant background appears in the s-wave channel with I=0, testing here the maximal possible interference.
However, this maximal interference scenario turns out to be unlikely since no comparable mass shifts have been observed in the spectral shape of the $\Sigma(1385)^+$ in the $\mathrm{\Lambda-p}$ final state \cite{Agakishiev:2011qw} or for the $\Lambda(1520)$ in the $\Sigma\pi$ decay channel. Moreover, it seems to be rather peculiar that interferences between the $\Lambda(1405)$ and the non$\--$resonant background should result in the same mass shift for both, the $\Sigma^+\pi^-$ and the $\Sigma^-\pi^+$ invariant mass distributions, where in both cases the physical origin of the non$\--$resonant background is quite different. According to \cite{Agakishiev:2012xk}, the $\Sigma^-\pi^+$ non$\--$resonant background arises from a strong contribution of a $\Delta^{++}$, whereas other mechanisms, e.g. $N^*/\Delta^0$ production via the reaction $p+p\rightarrow\Sigma^+ + N^*/\Delta^0+K^+\rightarrow\Sigma^++(\pi^-+p)+K^+$, could contribute to the $\Sigma^+\pi^-$ non$\--$resonant spectrum.\\
On the other hand, the presented results shall just emphasize that interference effects can play a significant role. Indeed, our model, even though it is not a full-fledged theoretical approach, has shown that interference effects can significantly shift the observed mass peak in the experimental spectra. We cannot prove that these effects are indeed responsible for the observed low mass $\Lambda(1405)$ signal. We just aim to illustrate the importance of a serious treatment of the non$\--$resonant background in any theoretical approach.
\section{Contributions of the two poles}
\begin{figure}[tbp]
	\centering
		\includegraphics[width=0.46\textwidth]{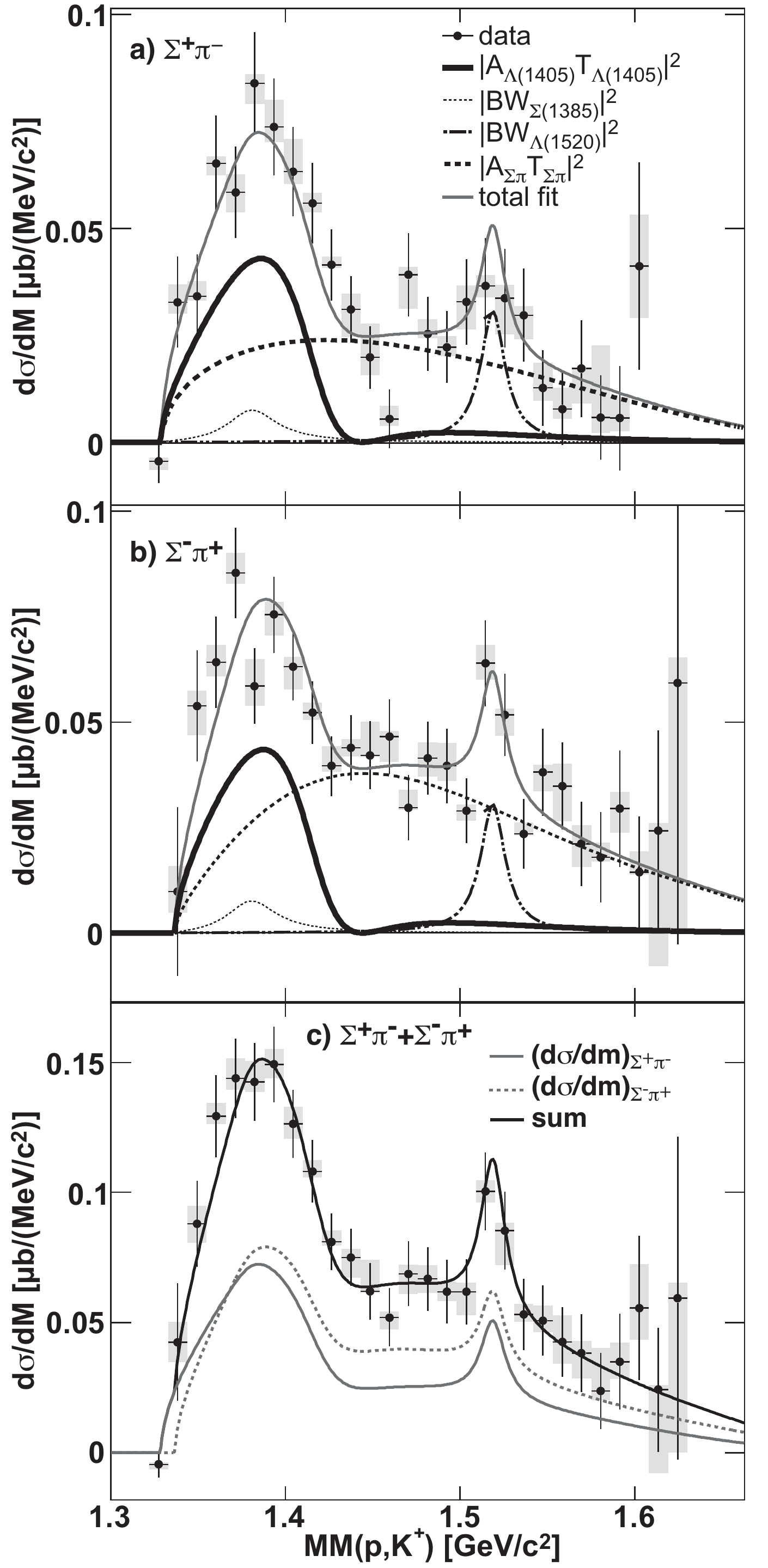}
	\caption{(Color online) Same as Fig.~\ref{fig:Interferences} but now the data are fitted with the Eq.s~(\ref{equ:DistSpPm2}) and (\ref{equ:DistSmPp2}). See text for details.}
	\label{fig:Fit_Weise}
\end{figure}

\begin{figure}[tbp]
	\centering
		\includegraphics[width=0.45\textwidth]{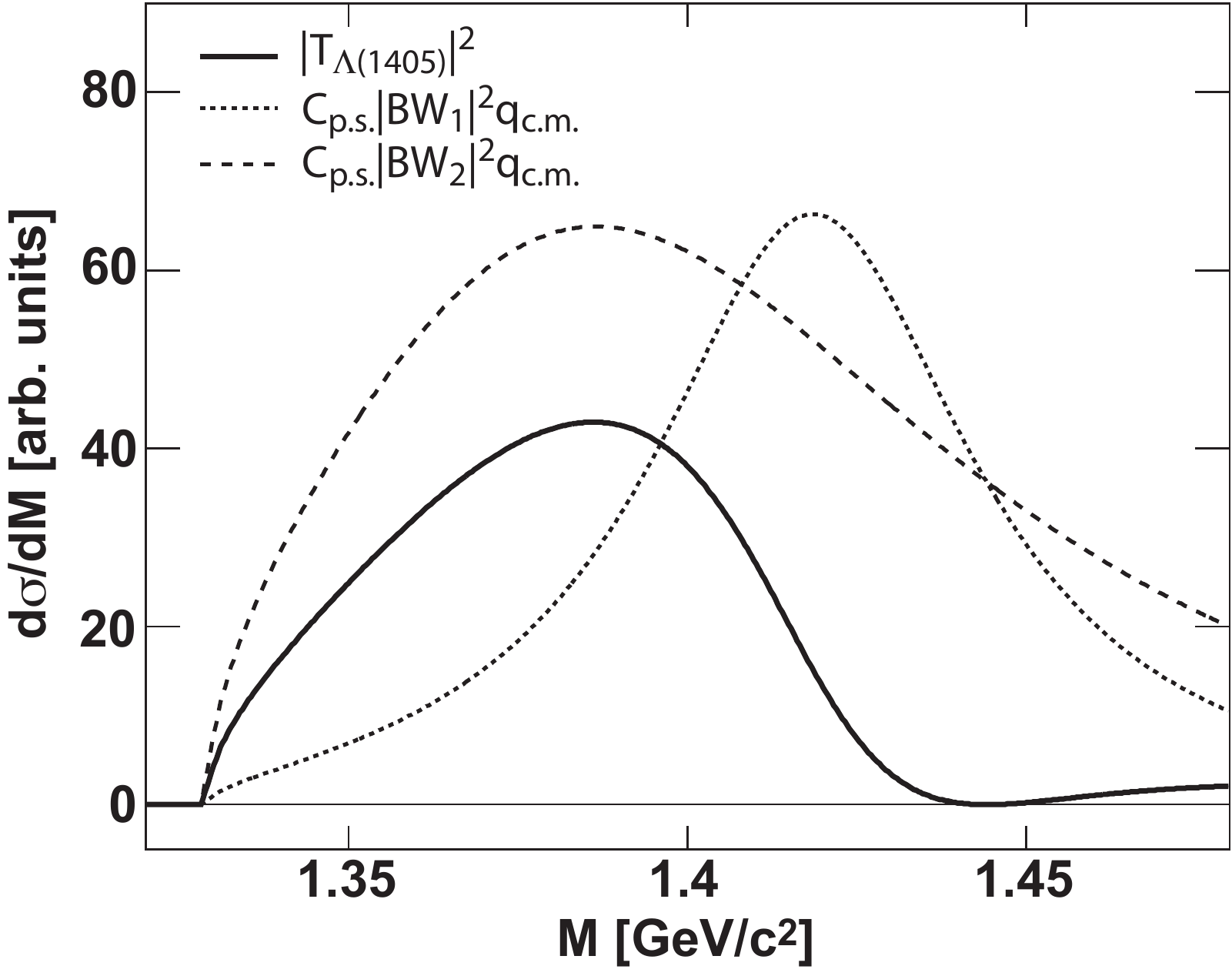}
	\caption{$\Lambda(1405)$ signal (black line) obtained by fitting Eq.~(\ref{equ:DistSpPm2}) and (\ref{equ:DistSmPp2}) to the experimental data in Fig.~\ref{fig:Fit_Weise}. The dotted and dashed lines show the contributions of $C_{p.s.}(m)\left|BW_1(m)\right|^2q_{c.m.}$ and $C_{p.s.}(m)\left|BW_2(m)\right|^2q_{c.m.}$.}
	\label{fig:L1405_Weise}
\end{figure}

\begin{figure}[tbp]
	\centering
		\includegraphics[width=0.46\textwidth]{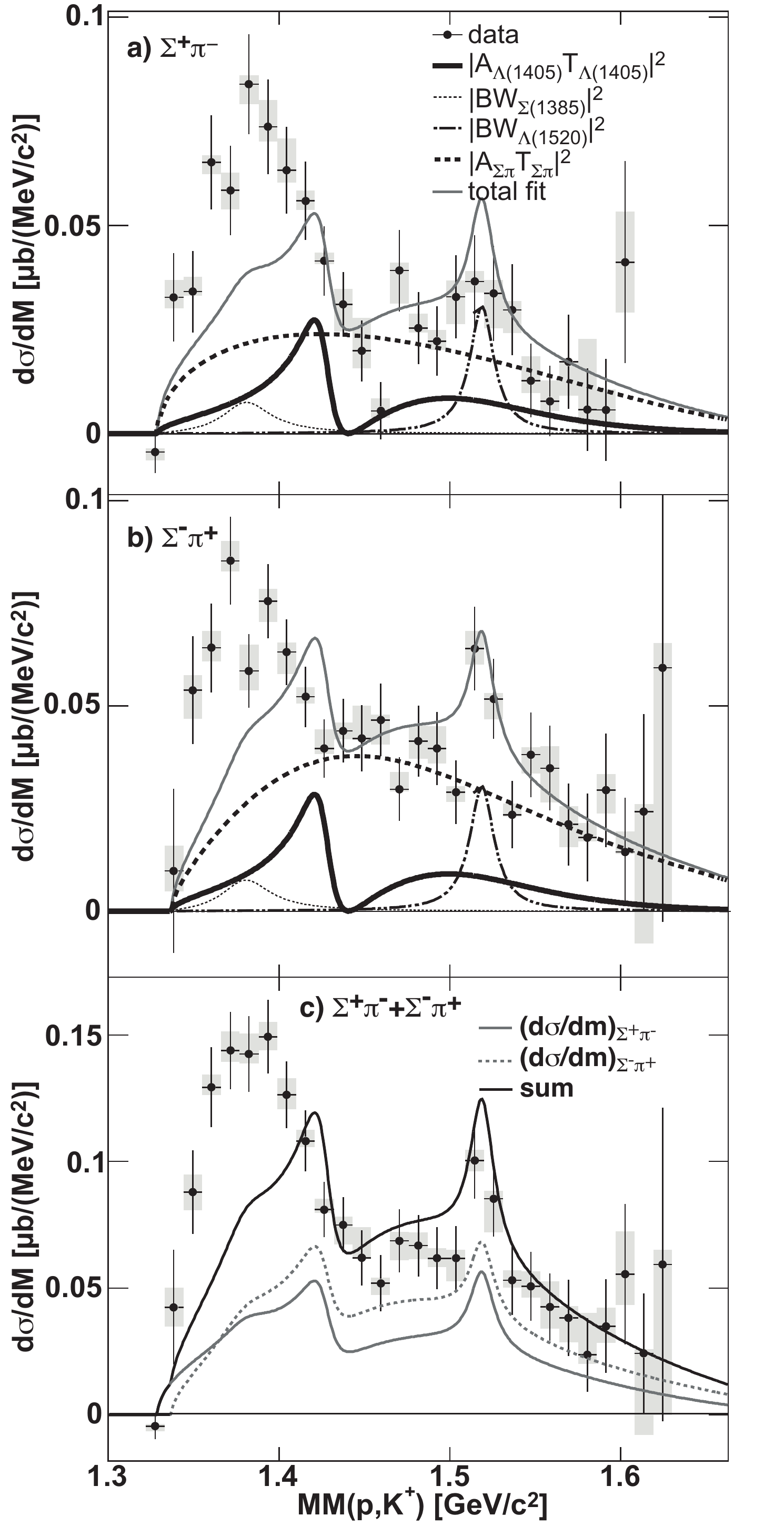}
	\caption{ Same as Fig.~\ref{fig:Fit_Weise} but the mass and width constraints for the two $\Lambda(1405)$ poles are taken from \cite{Mai:2012dt}. See text for details.}
	\label{fig:Fit_Meissner}
\end{figure}

\begin{figure}
	\centering
		\includegraphics[width=0.45\textwidth]{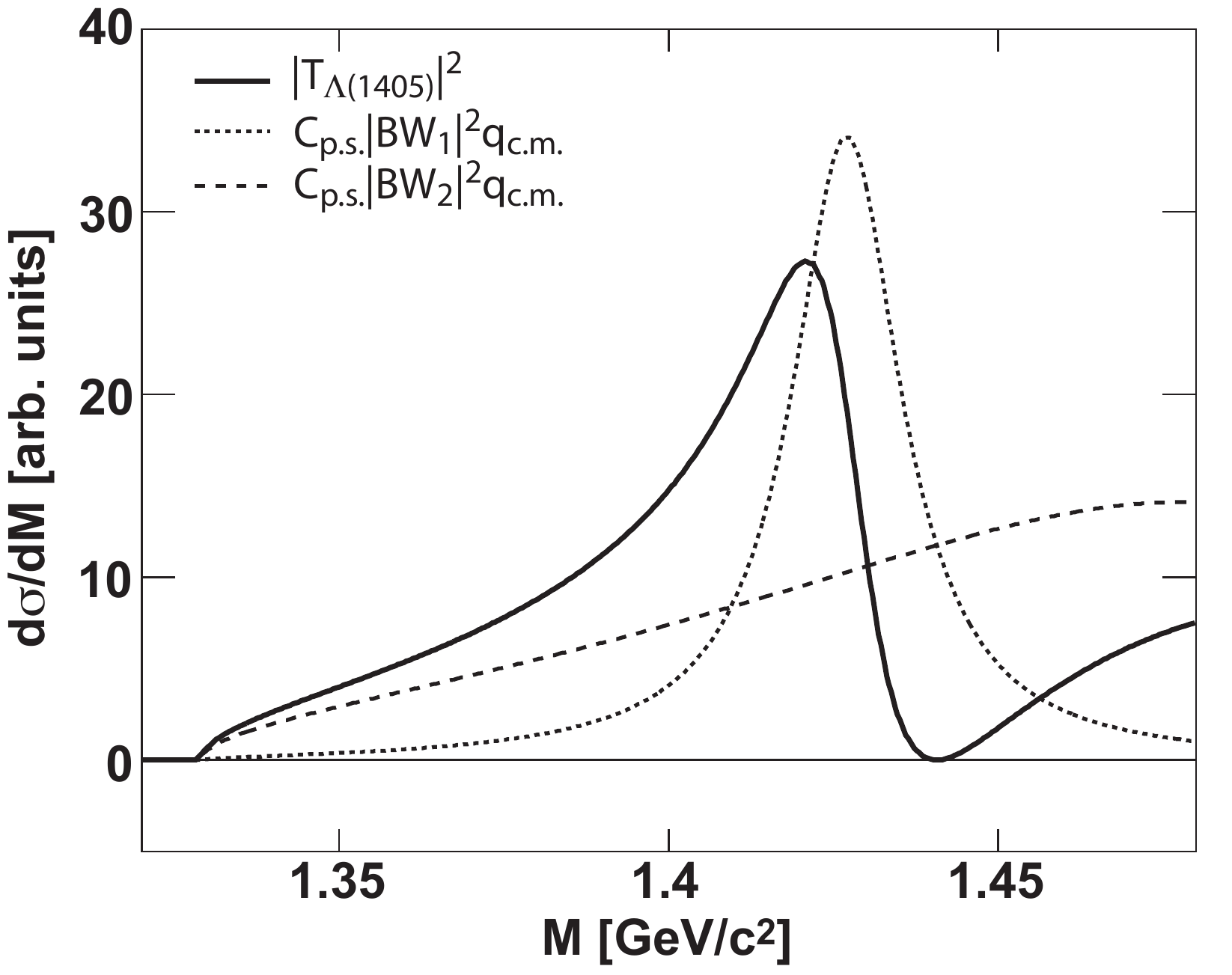}
	\caption{Same as Fig.~\ref{fig:L1405_Weise} but the constraints of \cite{Mai:2012dt} for the mass and width values of the two $\Lambda(1405)$ poles were used in the fits. See text for details.}
	\label{fig:L1405_Meissner1}
\end{figure}

Having illustrated the maximal possible influence of interference effects between the $\Lambda(1405)$ resonance and the non$\--$resonant background, we now consider the second extreme case, where this particular interference term is neglected. The observed low mass peaks in the HADES data are then completely attributed to the "pure'' $\Lambda(1405)$ signal. With this assumption one can try to determine the parameters and the relative contribution of the two $\Lambda(1405)$ poles. 
As a starting point, we again parameterize the $\Lambda(1405)$ as a coherent sum of the two Breit-Wigner amplitudes (see Eq.~(\ref{equ:DoubleBW})). This time, however, $m_{0,1}$, $\Gamma_{0,1}$, $m_{0,2}$ and $\Gamma_{0,2}$ as well as $A_1$ and $A_2$ shall be determined directly from the HADES data. As before, the real and imaginary part of both poles are constrained by the results of \cite{Ikeda:2012au} to $z_1=1424^{+7}_{-23}+i26^{+3}_{-14}$~MeV and $z_2=1381^{+18}_{-6}+i81^{+19}_{-8}$~MeV. Now the two functions~(\ref{equ:DistSpPm2}) and (\ref{equ:DistSmPp2}) are used to described the experimental data points of Fig.~\ref{fig:Interferences} a) and b). 
\begin{widetext}
\begin{eqnarray}
\left(\frac{d\sigma}{dm}\right)_{\Sigma^+\pi^-}=C_{p.s.}(m)\left|BW_1(m)e^{i\varphi_1}+BW_2(m) \right|^2q_{c.m.}+\left|A_{\Sigma^+\pi^-}T_{\Sigma^+\pi^-}\right|^2+\left|BW_{\Sigma(1385)^0}\right|^2 + \left|BW_{\Lambda(1520)}\right|^2 \label{equ:DistSpPm2} \\
\left(\frac{d\sigma}{dm}\right)_{\Sigma^-\pi^+}=C_{p.s.}(m)\left|BW_1(m)e^{i\varphi_1}+BW_2(m) \right|^2q_{c.m.}+\left|A_{\Sigma^-\pi^+}T_{\Sigma^-\pi^+}\right|^2+\left|BW_{\Sigma(1385)^0}\right|^2 + \left|BW_{\Lambda(1520)}\right|^2 \label{equ:DistSmPp2}
\end{eqnarray} 
\end{widetext}
In these Eq.s all individual contributions besides the two $\Lambda(1405)$ amplitudes sum up incoherently. The non$\--$resonant background is fixed in yield to the HADES results \cite{Agakishiev:2012xk}, by setting $A_{\Sigma^+\pi^-}$ and $A_{\Sigma^-\pi^+}$ to 1. In this way only $m_{0,1}$, $\Gamma_{0,1}$, $m_{0,2}$, $\Gamma_{0,2}$, $A_1$, $A_2$ and $\varphi_1$ are the free fit parameters in Eq.~(\ref{equ:DistSpPm2}) and (\ref{equ:DistSmPp2}). The results for the best fit ($\chi^2/ndf=1.04$) are shown in Fig.~\ref{fig:Fit_Weise} and the obtained fit parameters are listed in Table~\ref{tab:table3}.     
\begin{table}[tbp]
\caption{\label{tab:table3}%
Obtained free fit parameters after fit with Eq.s~(\ref{equ:DistSpPm2}) and (\ref{equ:DistSmPp2}).}
\begin{ruledtabular}
\begin{tabular}{cccccc}
\textrm{$m_{0,1}$}&
\textrm{$\Gamma_{0,1}$}&
\textrm{$m_{0,2}$}&
\textrm{$\Gamma_{0,2}$}&
\textrm{$A_1/A_2$}&
\textrm{$\varphi_1$}\\
\colrule
$1418$ & $58$ & $1375$ & $146$ & $0.395$ & $178^{\circ}$ \\
\end{tabular}
\end{ruledtabular}
\end{table}
A very good description of the data is achieved with the maximum in the $\Lambda(1405)$ distribution now appearing at around $1385$~MeV/$c^2$. Integrating this signal results in a total cross section of $\sigma=9.0$~$\mu b$ for the reaction $p+p\rightarrow\Lambda(1405)+p+K^+$, which is in good agreement with the result quoted in \cite{Agakishiev:2012xk}. The composition of the $\Lambda(1405)$ signal itself is illustrated in Fig.~\ref{fig:L1405_Weise}. The resulting $\Lambda(1405)$ amplitude differs strongly from the results reported in \cite{Geng:2007vm} and shown in Fig. \ref{fig:OsetFitted}. 
The $z_2$ pole (dashed line) has the dominant contribution with mass and width values at the edge of the allowed fit range ($m_{0,2}=1375$~MeV, $\Gamma_{0,2}=146$~MeV). This very broad signal interferes with the $z_1$ pole (dotted line) in a way that the high mass region of the $\Lambda(1405)$ is strongly suppressed, creating a peak below 1400~MeV/$c^2$ (solid black line). However, with the large amount of free parameters, with the broad ranges in which the mass and width values are constrained and with the limited number of data points, it is difficult to derive strict conclusions about the relative contributions and the positions of the two poles. It can just be claimed that a significant part of the $\Lambda(1405)$ amplitude could be associated to the second, broad pole. One should also notice here that the $\Lambda(1405)$ signal shows a tail for masses above 1440~MeV/$c^2$. This tail was not considered by the authors of \cite{Agakishiev:2012xk}, who assumed the $\Lambda(1405)$ to be only located at lower energies and who therefore used the high mass range of the $\Sigma\pi$ spectra to determine the contribution of the $\Sigma\pi$ non$\--$resonant channels. From the results in Fig.~\ref{fig:Fit_Weise} and \ref{fig:L1405_Weise} one sees, however, that this assumption might be too simple. Thus, a further possibility is to not fix the non$\--$resonant background but to treat $A_{\Sigma^+\pi^-}$ and $A_{\Sigma^-\pi^+}$ as two additional fit parameters.
Applying this new fit to the data, results in the fit values listed in Table~\ref{tab:table4}.
\begin{table}[tbp]
\caption{\label{tab:table4}%
Obtained free fit parameters after fit with Eq.s~(\ref{equ:DistSpPm2}) and (\ref{equ:DistSmPp2}) with $A_{\Sigma^+\pi^-}$ and $A_{\Sigma^-\pi^+}$ as additional free parameters.}
\begin{ruledtabular}
\begin{tabular}{cccccccc}
\textrm{$m_{0,1}$}&
\textrm{$\Gamma_{0,1}$}&
\textrm{$m_{0,2}$}&
\textrm{$\Gamma_{0,2}$}&
\textrm{$A_1/A_2$}&
\textrm{$\varphi_1$} &
\textrm{$A_{\Sigma^+\pi^-}$} & 
\textrm{$A_{\Sigma^-\pi^+}$} \\
\colrule
$1431$ & $58$ & $1375$ & $146$ & $0.204$ & $164^{\circ}$ & $0.857$ & $0.887$\\
\end{tabular}
\end{ruledtabular}
\end{table} 
The fit quality of $\chi^2/ndf=1.03$ is as good as before, but the contribution of the $z_1$ pole is further reduced as seen by comparing the $A_1/A_2$ ratios in Table~\ref{tab:table3} and \ref{tab:table4}. However, as the number of free parameters has further increased, the fit to the data is not robust anymore so that the results of Table~\ref{tab:table4} are not very reliable.\\ 
In summary, it can be concluded that it is rather difficult to precisely determine the relative contribution between $z_1$ and $z_2$ and simultaneously to determine their exact positions in the complex energy plane just by a fit to the HADES data alone. In fact, an appropriate theory model for proton-proton reactions with a serious treatment of all possible background contributions is required to make further conclusions. Nevertheless, the results obtained in this work clearly show that, in order to describe the new HADES data, a rather large contribution of the broad, low mass $\Lambda(1405)$ pole is needed, provided that no interference with the non$\--$resonant background is present.\\
In this context, it would also be important to have more restrictive constraints on the real and imaginary part of $z_1$ and $z_2$. The values quoted above allow a rather large variation of these parameters. However, the situation is even more unclear. In a recent work by Mai and Meissner \cite{Mai:2012dt}, who also used the latest SIDDARTHA results, the positions of the two poles were derived to $z_1=1428^{+2}_{-1}+i8^{+2}_{-2}$~MeV and $z_2=1497^{+11}_{-7}+i75^{+9}_{-9}$~MeV. 
The imaginary part of the first pole is much smaller than the one of \cite{Ikeda:2012au}, but even more spectacular is the totally different value in the real part of $z_2$, which is shifted by about 100~MeV to higher energies. We can also take these values to fit the HADES data. For this purpose, the non$\--$resonant background amplitudes $A_{\Sigma^+\pi^-}$ and $A_{\Sigma^-\pi^+}$ are again fixed to 1 and the mass and width values are allowed to vary within the given ranges. The best fit result is shown in Fig.~\ref{fig:Fit_Meissner} and the obtained fit parameters are listed in Table~\ref{tab:tableMeissner}. The corresponding decomposition of the $\Lambda(1405)$ spectrum is shown in Fig.~\ref{fig:L1405_Meissner1}. 

\begin{table}[tbp]
\caption{\label{tab:tableMeissner}%
Obtained free fit parameters after fit with Eq.s~(\ref{equ:DistSpPm2}) and (\ref{equ:DistSmPp2}) and by using the constraints of \cite{Mai:2012dt} for the mass and width values of the two $\Lambda(1405)$ poles.}
\begin{ruledtabular}
\begin{tabular}{cccccc}
\textrm{$m_{0,1}$}&
\textrm{$\Gamma_{0,1}$}&
\textrm{$m_{0,2}$}&
\textrm{$\Gamma_{0,2}$}&
\textrm{$A_1/A_2$}&
\textrm{$\varphi_1$}\\
\colrule
$1427$ & $20$ & $1490$ & $168$ & $0.17$ & $264^{\circ}$ \\
\end{tabular}
\end{ruledtabular}
\end{table}      
The fit result is very poor, which is also expressed in a normalized $\chi^2$-value of $\chi^2/ndf=3.4$. With the $z_2$ pole having such a large real part, it becomes impossible to create a peak structure at around 1385~MeV/$c^2$ like it was observed by the HADES collaboration. Within our simplified model, we can thus conclude that the pole positions extracted in \cite{Mai:2012dt} are not compatible with the new HADES data.

\section{Summary} 

We have presented different interpretations of the low mass $\Lambda(1405)$ signal, measured by the HADES collaboration in $p+p$ reactions. It was shown that the obtained signal is very similar to the results obtained in $\pi^-+p$ reactions. One possible explanation for the observed mass shift is based on interference effects. For that purpose we have developed a simple model, where we assumed the $\Lambda(1405)$ to consist of two poles, parameterized as Breit-Wigner amplitudes. The contribution and position of these poles were first chosen such to match the  $\Lambda(1405)$ line shape of \cite{Geng:2007vm} with the mass peak appearing at $\approx 1410$~MeV/$c^2$. With this model we could show that maximum interference between the $\Lambda(1405)$ amplitude and the non$\--$resonant $\Sigma\pi$ background can indeed create a low mass signal and thus describe the HADES data. However, it was argued that this explanation is probably very unrealistic. In a second approach we have neglected interference effects with the non$\--$resonant background and have used the HADES data themselves to determine the position and the relative contribution of the two $\Lambda(1405)$ poles. With a relatively large contribution of the second, broad $\Lambda(1405)$ pole, $z_2$, we could achieve a very good description of the HADES data. However, the limited statistic in the data did not allow to draw firm conclusions on the precise contribution of the two poles. Also their position in the complex energy plane could not be determined precisely just by a fit to this single data set. Nevertheless, it was shown that, within our simple model, a high energy pole position of $z_2=1497^{+11}_{-7}+i75^{+9}_{-9}$~MeV, like it was obtained in \cite{Mai:2012dt}, is not compatible with the new HADES data. \\

The authors would like to thank Wolfram Weise, Wolfgang Koenig and Piotr Salabura for very useful discussions. This work is supported by the Munich funding agency, BMBF 06DR9059D,05P12WOGHH, TU M\"unchen, Garching
(Germany) MLL M\"unchen: DFG EClust 153, VH-NG-330 BMBF 06MT9156 TP6 GSI.     

\bibliography{L1405PaperInterV1}

\end{document}